\documentclass[showpacs,12pt,preprintnumbers,amssymb]{revtex4}
\usepackage{graphicx}
\usepackage[english]{babel}

\begin{document}

\bibliographystyle{apsrev}

\title{Memory functions and Correlations in  Additive Binary Markov Chains}

\author{S. S. Melnyk, O. V. Usatenko, V. A. Yampol'skii
\footnote[1]{yam@ire.kharkov.ua} }
\affiliation{A. Ya. Usikov Institute for Radiophysics and Electronics \\
Ukrainian Academy of Science, 12 Proskura Street, 61085 Kharkov,
Ukraine}

\author{S. S. Apostolov, Z.A. Maiselis}
\affiliation{V. N. Karazin Kharkov National University, 4 Svoboda
Sq., Kharkov 61077, Ukraine}

\begin{abstract}
A theory of additive Markov chains with long-range memory, proposed
earlier in Phys. Rev. E \textbf{68}, 06117 (2003), is developed and
used to describe statistical properties of long-range correlated
systems. The convenient characteristics of such systems, a memory
function, and its relation to the correlation properties of the
systems are examined. Various methods for finding the memory
function via the correlation function are proposed. The inverse
problem (calculation of the correlation function by means of the
prescribed memory function) is also solved. This is demonstrated for
the analytically solvable model of the system with a step-wise
memory function.
\end{abstract}
\pacs{05.40.-a, 02.50.Ga, 87.10.+e}

\maketitle

\section{Introduction}
\label{I}

 The problem of long-range correlated dynamic systems (LRCS)
has been under study for a long time in many areas of contemporary
physics~\cite{bul,sok,bun,yan,maj,halvin},
biology~\cite{vossDNA,stan,buld,prov,yul,hao},
economics~\cite{stan,mant,zhang},
literature~\cite{schen,kant,kokol,ebeling,uyakm},
etc.~\cite{stan,czir}. One of the ways to get a correct insight into
the nature of correlations in a system consists in constructing a
mathematical object (for example, a correlated sequence of symbols)
possessing the same statistical properties as the initial system.
There exist many algorithms for generating long-range correlated
sequences: the inverse Fourier transformation~\cite{czir,maks}, the
expansion-modification Li method~\cite{li}, the Voss procedure of
consequent random additions~\cite{voss}, the correlated Levy
walks~\cite{shl}, etc.~\cite{czir}. The use of the \emph{multi-step
Markov} chains is one of the most important among them because they
offer a possibility to construct a random sequence with necessary
correlated properties in the most natural way. This was demonstrated
in Ref.~\cite{uya}, where the concept of Markov chain with the
\emph{step-wise memory function} was introduced. The correlation
properties of some dynamical systems (coarse-grained sequences of
the \emph{Eukarya's DNA and dictionaries}) can be well described by
this model~\cite{uya}.

A sequence of symbols in the Markov chain can be thought of as the
sequence of states of certain particle, which participates in a
correlated Brownian motion. Thus, every $L$-word (the portion of the
length $L$ in the sequence) can be considered as one of the
realizations of the ensemble of correlated Brownian trajectories in
the "time" interval $L$. This point gives an opportunity to use the
statistical methods for examining the correlation properties of the
dynamic systems. Another important reason for the study of Markov
chains is its application to the various physical
objects~\cite{tsal,abe,den}, e.g., to the Ising chains of spins. The
problem of thermodynamics description of the Ising chains with
long-range spin interaction is still unresolved even for the 1D
case. However, the association of such systems with the Markov
chains can shed light on the non-extensive thermodynamics of the
LRCS.

In this paper, we ascertain the relation between the memory function
of the additive Markov chains and the correlation properties of the
systems under consideration. We examine the simplest  variant of the
random sequences, dichotomic (binary) ones, although the presented
theory can be applied to arbitrary additive Markov processes with
finite or infinite number of states.

The paper is organized as follows. In the first Section, we
introduce the general relations for the Markov chains, derive an
equation connecting the correlation and memory functions of additive
Markov chains, and verify the robustness of our method by numerical
simulations. The second part is devoted to the study of the
correlation function for the Markov chain with the step-wise memory
function. In Subsec.~(\ref{CF}) we reveal a band structure of the
correlation function and obtain its explicit expression.
Subsec.~(\ref{AS}) contains the results of asymptotic study of
correlation function.

\section{General Properties of Additive Markov Chains}
\label{GP}
\subsection{Basic notions}
\label{BN}

 Let us consider a homogeneous binary sequence of symbols,
$a_{i}=\{0,1\}$, $i\in\textbf{\textbf{Z}} =...,-2,-1,0,1,2,...$. To
determine the $N$-\textit{step Markov chain} we have to introduce
the \emph{conditional probability} $P(a_{i}\mid
a_{i-N},a_{i-N+1},\dots ,a_{i-1})$ of  the definite symbol $a_i$
(for example, $a_i =1$ or $a_i =0$ ) occurring after the $N$-word
$T_{N,i}$, where $T_{N,i}$ denotes the sequence of symbols
$a_{i-N},a_{i-N+1},\dots ,a_{i-1}$. Thus, it is necessary to define
$2^{N}$ values of the $P$-function corresponding to each possible
configuration of the symbols in the $N$-word $T_{N,i}$. Since we
intend to deal with the sequences possessing the memory length of
the order of $10^6$, we need to make some simplifications. We
suppose that the $P$-function has the \textit{additive} form,
\begin{equation}\label{mark1}
P(a_{i}=1\mid T_{N,i}) = \sum\limits_{k=1}^{N}f(a_{i-k},k).
\end{equation}
Here the value $f(a_{i-k},k)$ is the additive contribution of the
symbol $a_{i-k}$ to the conditional probability of the symbol unity
occurring at the $i$th site. Equation~(\ref{mark1}) corresponds to
the additive influence of the previous symbols on the generated one.
Such Markov chain is referred to as \emph{additive Markov chain},
Ref.~\cite{mel}. The homogeneity of the Markov chain is provided by
the independence of the conditional probability Eq.~(\ref{mark1}) of
the index $i$. It is possible to consider Eq.~(\ref{mark1}) as the
first term in expansion of conditional probability in the formal
series of terms that correspond to the additive (or unary), binary,
ternary, and so on functions up to $N$-ary one.

Let us rewrite Eq.~(\ref{mark1}) in an equivalent form,
\begin{equation}\label{MF}
P(a_{i}=1\mid T_{N,i})=\bar{a}+\sum\limits_{r=1}^{N}
F(r)(a_{i-r}-\bar{a}).
\end{equation}
Here
\[
\bar{a}=\sum\limits_{r=1}^{N}f(0,r)/[1-\sum\limits_{r=1}^{N}
(f(1,r)-f(0,r))]
\]
is the average number of unities in the sequence, Ref.~\cite{mel},
and
\[
F(r)=f(1,r)-f(0,r).
\]
We refer to $F(r)$ as the \emph{memory function} (MF). It describes
the strength of impact of previous symbol $a_{i-r}$ upon a generated
one, $a_{i}$. Evidently, this function has to satisfy condition
$0\leqslant P(a_{i}=1\mid T_{N,i})\leqslant 1$. To the best of our
knowledge, the concept of the memory function for multi-step Markov
chains was introduced in papers~\cite{uyakm,uya} where it is shown
that it is convenient to use it in describing the correlated
properties of complex dynamical systems with long-range
correlations.

The function $P(a_{i}=1\mid T_{N,i})$ contains complete information
about correlation properties of the Markov chain. In general, the
correlation function and other moments are employed as the input
characteristics for the description of the correlated random
systems. Yet, the correlation function takes account of not only the
direct interconnection of the elements $a_i$ and $a_{i+r}$, but also
their indirect interaction via other elements. Our approach operates
with the ``origin'' characteristics of the system, specifically with
the memory function.

The positive values of the MF result in persistent diffusion where
previous displacements of the Brownian particle in some direction
provoke its consequent displacement in the same direction. The
negative values of the MF correspond to the antipersistent diffusion
where the changes in the direction of motion are more probable. In
terms of the Ising long-range particles interaction model, which
could be naturally associated with the Markov chains, the positive
values of the MF correspond to the attraction of particles and
negative ones conform to the repulsion.

We consider the distribution $W_{L}(k)$ of the words of definite
length $L$ by the number $k$ of unities in them,
$k_{i}(L)=\sum\limits_{l=1}^{L}a_{i+l}$, and the variance $D(L)$
of $k_{i}(L)$,
\begin{equation}\label{defD}
D(L)=\overline{(k-\bar{k})^{2}},
\end{equation}
where the definition of average value of $g(k)$ is
$\overline{g(k)}=\sum\limits_{k=0}^{L}g(k)W_{L}(k)$.

Another statistical characteristics of random sequences is the
correlation function,
\begin{equation}\label{defK}
K(r)=\overline{a_{i}a_{i+r}}-\bar{a}^{2}.
\end{equation}
By definition, the correlation function is even, $K(-r)=K(r)$, and
$K(0)=\bar{a}(1-\bar{a})$ is the variance of the random variable
$a_i$. The correlation function is connected with the above
mentioned variance by the equation
\begin{equation}\label{KD}
K(r)=\frac{1}{2}(D(r-1)-2D(r)+D(r+1)),
\end{equation}
or
\begin{equation}\label{mark4}
K(r)=\frac{1}{2}\frac{\textrm{d}^2D(r)}{\textrm{d}r^2}
\end{equation}
in the continuous limit.

\subsection{Derivation of main equation}
\label{ME}

 In this subsection we obtain a very important relation
connacting the memory and correlation functions of the additive
Markov chain. Let us introduce the function $\phi(r)=p(a_{i}=1\mid
a_{i-r}=1)$, which is the probability of symbol $a_{i}=1$ occurring
under condition that previous symbol $a_{i-r}$ is also equal to
unity. This function is obviously connected to the correlation
function $K(r)$, see Eq.~(\ref{defK}) since the quantity
$\overline{a_{i}a_{i-r}}$ is the probability of simultaneous
equality to unity of both symbols, $a_{i}$ and $a_{i-r}$. It can be
expressed in terms of the conditional probability $\phi(r)$,
\begin{equation}\label{main4}
\overline{a_{i}a_{i-r}}=\bar{a}\phi(r).
\end{equation}
Substituting Eq.~(\ref{MF}) into Eq.~(\ref{defK}) we get:
\begin{equation}\label{main5}
K(r)=\bar{a}\phi(r)-\bar{a}^{2}.
\end{equation}

For the $N$-step Markov chain, the probability of the symbol
$a_{i}=1$ occurring depends only on the previous $N$-word.
Therefore, to obtain the value of $\phi(r)$ one needs to average
the conditional probability $P$ Eq.~(\ref{MF}) over all
realizations of the $N$-words with the condition $a_{i-r}=1$,
\begin{equation}\label{main6}
\phi(r)=p(a_{i}=1\mid a_{i-r}=1)
\end{equation}
\[
=\sum\limits_{T_{N,i}}P(a_{i}=1\mid T_{N,i})p(T_{N,i}\mid
a_{i-r}=1).
\]
If the value of $r$ is less or equal to $N$, then $a_{i-r}$ in
Eq.~(\ref{main6}) is one of the symbols $a_{i-1},a_{i-2},\dots
,a_{i-N}$ in the word $T_{N,i}$. In this case, the sum in
Eq.~(\ref{main6}) allows not all $N$-words, but only the words that
contain the symbol unity at the $(i-r)$th position. If $r>N$, the
memory function $F(r)$ equals zero in this region and, therefore,
the sum in Eq.~(\ref{main6}) contains \emph{all} terms corresponding
to all different $N$-words.

Substituting Eq.~(\ref{MF}) into Eq.~(\ref{main6}), we obtain:
\begin{equation}\label{main7}
\phi(r)=\bar{a}\sum\limits_{T_{N,i}}P(T_{N,i}\mid a_{i-r}=1)
\end{equation}
\[
+\sum\limits_{r'=1}^{N}F(r')\sum\limits_{T_{N,i}}(a_{i-r'}-\bar{a})P(T_{N,i}\mid
a_{i-r}=1).
\]

According to the normalization condition, the first sum in
Eq.~(\ref{main7}) is equal to unity. Consider the sum
\begin{equation}\label{main8}
\sum\limits_{T_{N,i}}a_{i-r'}P(T_{N,i}\mid a_{i-r}=1)
\end{equation}
in the second term in RHS of Eq.~(\ref{main7}). The symbol
$a_{i-r'}$ is contained within the word $T_{N,i}$. Therefore,
Eq.~(\ref{main8}) represents the average value of $a_{i-r'}$ under
condition $a_{i-r}=1$. In other words, it equals the probability
$\phi(r-r')$ of $a_{i-r'}=1$ occurring under the condition
$a_{i-r}=1$:
\begin{equation}\label{main9}
\sum\limits_{w}a_{i-r'}P(T_{N,i}\mid a_{i-r}=1)=\phi(r-r').
\end{equation}
Substituting this equation into Eq.~(\ref{main7}) we obtain:
\begin{equation}\label{main10}
\phi(r)=\bar{a}+\sum\limits_{r'=1}^{N}F(r')(\phi(r-r')-\bar{a}).
\end{equation}
Taking into account Eq.~(\ref{main5}), we arrive at the relation
between the memory function and the correlation function:
\begin{equation}\label{main}
K(r)=\sum\limits_{r'=1}^{N}F(r')K(r-r'),\qquad r\geqslant 1.
\end{equation}
This equation was first derived by the variation method in
Ref.~\cite{mel}.

Another equation resulting from Eq.~(\ref{main}) by double summation
over index $r$ establishes a relationship between the memory
function $F(r)$ and the variance $D(L)$,
\begin{equation}\label{mainD}
M(r,0)=\sum\limits_{r'=1}^{N}F(r')M(r,r'),
\end{equation}
\[
M(r,r')=D(r-r')-(D(-r')+r[D(-r'+1)-D(-r')]).
\]
Equation~(\ref{KD}) and parity of the function $D(r)$ are used here.

The last equation shows convenience of using the variance $D(L)$
instead of the correlation function $K(r)$. The function $K(r)$,
being a second derivative of $D(r)$ in continuous approximation, is
less robust in computer simulations. It is the main reason why we
prefer to use Eq.~(\ref{mainD}) for the long-range memory sequences.
This is our tool for finding the memory function $F(r)$ of a
sequence using the variance $D(L)$.

\subsection{Numerical reconstruction of the memory function }
\label{NS}

Let us verify the robustness of our method by numerical
simulations. We consider a model \emph{memory function},
\begin{equation}
F(r)=0.1\cases {1-r/10,\;\qquad \;\;\ 1 \leqslant r < 10,  \cr 0,
\;\;\;\;\;\;\;\;\;\;\ \qquad \qquad r\geqslant 10.} \label{eqmf}
\end{equation}
\protect\begin{figure}[h!]
\begin{centering}
\scalebox{0.8}[0.8]{\includegraphics{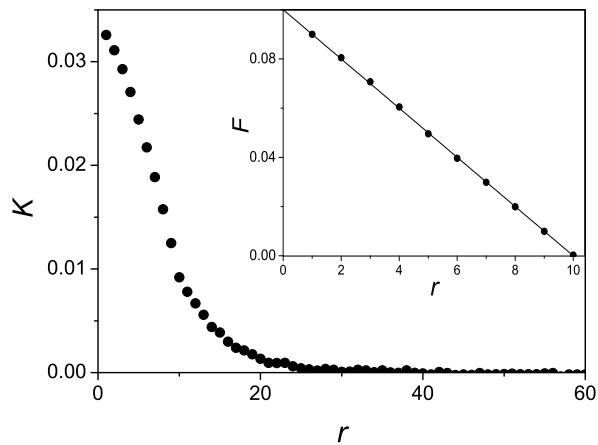}} \caption{Calculated
correlation function $K(r)$ of the Markov chain constructed with the
model memory function $F(r)$, Eq.~(\ref{eqmf}), shown by the solid
line in the inset. Dots in the inset correspond to the memory
function reconstructed by solving Eq.~(\ref{main}) with correlation
function $K(r)$ presented in the main panel.} \label{Line}
\end{centering}
\end{figure}
shown in the inset in Fig.~\ref{Line} by solid line. Using
Eq.~(\ref{MF}), we construct a random unbiased $\bar{a}=1/2$, Markov
chain. Then, with the help of the constructed binary sequence of the
length $3\times10^6$, we calculate numerically the correlation
function $K(r)$ by solving the set of $N$ linear
equations~(\ref{main}). The result of these calculations is given in
Fig.~\ref{Line}. One can see that the correlation function $K(r)$
mimics roughly the memory function $F(r)$ over the region
$1\leqslant r \leqslant 10$. In the region $r>10$, the memory
function is equal to zero but the correlation function does not
vanish~\cite{ref}. Then, using the obtained correlation function
$K(r)$, we solve numerically Eq.~(\ref{main}). The result is shown
in the inset in Fig.~\ref{Line} by dots. One can see an excellent
agreement of initial, Eq.~(\ref{eqmf}), and reconstructed memory
functions $F(r)$.

The ability of constructing a binary sequence with an arbitrary
\emph{prescribed correlation function} by means of Eq.~(\ref{main})
is the very nontrivial result of this paper.

Yet another approach to numerical finding the memory function is an
iteration procedure. For its realization let us rewrite
Eq.~(\ref{main}) in the form,
\begin{equation}
\label{iter0} F(r)=\frac {K(r)}{K(0)}-\!\!\!\sum\limits_{r'=1,\
r'\neq r}^N\!\!\!\frac {K(r-r')}{K(0)}F(r').
\end{equation}
Using Eq.~(\ref{iter0}) with starting iteration $F_0(r)=0$, we
obtain the formula,
\begin{equation}\label{iter1}
F_{n+1}(r)=\frac {K(r)}{K(0)}-\!\!\!\sum\limits_{r'=1,\ r'\neq
r}^N\!\!\!\frac {K(r-r')}{K(0)}F_n(r'),\,n\geqslant0.
\end{equation}
Thus, the memory function can be presented as the series,
\begin{equation}\label{iter}
F(r)=\frac {K(r)}{K(0)}-\sum\limits_{r'\neq r}\frac
{K(r-r')K(r')}{K^2(0)}+\sum\limits_{r'\neq r}\sum\limits_{r''\neq
r'}\frac {K(r-r')K(r'-r'')K(r'')}{K^3(0)}+\dots
\end{equation}
Note, that the Markov chain with the definite correlation function
$K(r)$ exists if the series~(\ref{iter}) is convergent and the
obtained function implies the probability Eq.~(\ref{MF})
satisfying the requirement $0 \leqslant P(a_{i}=1\mid
T_{N,i})\leqslant 1$ for arbitrary word $T_{N,i}$. If
$\bar{a}=1/2$ we obtain the restriction, $\sum |F(r)| \leqslant
1$. The sufficient, but not necessary, requirement is
$\sum\limits_{r=1}^{N-1} |K(r)|\leqslant 1/12-|K(N)|/3$.

\section{CORRELATION FUNCTION of the chain with the step-wise memory function}
\label{CF-SW}

In the previous section, we obtained the relationship (\ref{main})
between two characteristics of the Markov chain, the memory and
correlation functions, and used this equation to solve the problem
of finding the memory function via the known correlation function.
Here we present the solution of the inverse problem. We suppose the
memory function to be known and find the correlation function of the
corespondent Markov chain. To simplify our consideration, we examine
the step-wise memory function,
\begin{equation}\label{mfsw}
F(r)=\cases{\alpha,\quad r\leqslant N, \cr 0,\quad r>N.}
\end{equation}
The restriction imposed on the parameter $\alpha$ can be obtained
from Eq.~(\ref{MF}): $|\alpha|<1/N$. Note that each of the symbols
unity in the preceding N-word promotes the emergence of new symbol
unity if $0 < \alpha <1/N$. This corresponds to the persistent
diffusion. The region of parameter $\alpha$, determined by
inequality $-1/N<\alpha <0$, corresponds to the anti-persistent
diffusion. If $\alpha=0$, one arrives at the case of the
non-correlated Brownian motion.

\subsection{Main equation for the correlation function}
\label{MECF}

Substituting Eq.~(\ref{mfsw}) into Eq.~(\ref{main}) we obtain the
relation,
\begin{equation}\label{base}
K(r)=\alpha\sum\limits_{r'=1}^{N} K(r-r'),\quad r\geqslant 1.
\end{equation}
Here, the correlation function is assumed to be even, $K(-r)=K(r)$.
Equation (\ref{base}) is the linear recurrence of the order of $N$
for $r\geqslant N+1$, so we stand in need of $N$ initial conditions.
For the unbiased sequence, $\bar{a}=1/2$, we have $K(0)=1/4$. The
solution of Eqs.~(\ref{base}) written for $r=1, \dots , N$ yields
the constant value of the correlation function, $K(r)=K_0$ at $r=1,
\dots ,N-1$,
\begin{equation}\label{b}
K_0=\frac{\alpha}{4(1-\alpha(N-1))}.
\end{equation}

Subtracting Eq.~(\ref{base}) from the same equation written for
$r+1$, we derive another, more convenient, form of the recurrence:
\begin{equation}\label{basem}
K(r+1)-(1+\alpha)~ K(r)+\alpha ~ K(r-N)=0.
\end{equation}
This equation is of the order of $N+1$, thus we need an additional
initial condition. It can be derived from Eq.~(\ref{base}):
$K(N)=K_0$.  Note that the possibility to rewrite Eq.~(\ref{base})
in the form of Eq.~(\ref{basem}) is the result of the simple
structure of the memory function. We solve the obtained recurrent
equations by the most natural method, by means of step-by-step
finding the sequent values of the correlation function. Such an
approach is very suitable for the analysis of correlation function
at $r\gtrsim N$.

\subsection{Correlation function at $r\gtrsim N$}
\label{CF}

\subsubsection{Band structure of the correlation function}
\label{ZS}

\begin{figure}[h!]
\begin{centering}
\scalebox{0.55}[0.5]{\includegraphics{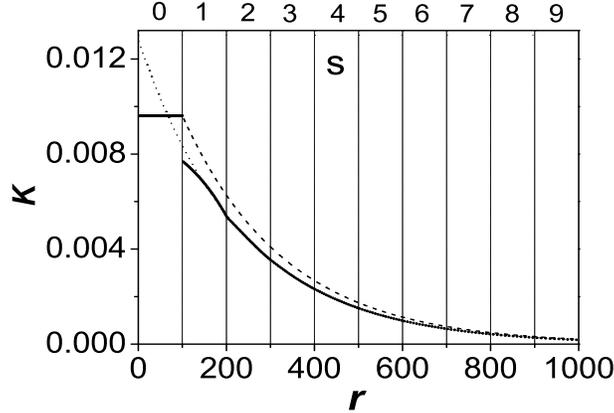}}
\caption{Correlation function $K(r)$ (solid line) obtained by two
different methods: the numerical simulation of Eq.~(\ref{main}) and
exact solution Eq.~(\ref{solz}). The dotted line is for the
contribution to $K(r)$ of the first root of Eq.~(\ref{solM-0}). The
dashed line refers to the correlation function obtained in
\cite{uyakm}. The vertical lines indicate the limits of the bands
numbered by $s$.} \label{korf}
\end{centering}
\end{figure}

Equation~(\ref{base}) allows one to find numerically the unknown
correlation function $K(r)$. The result of this step-by-step
calculation is presented in Fig.~\ref{korf} by solid line.
\protect
One can easily see the discontinuity of $K(r)$ at the point
$L=N=100$, the breakpoint of the curve is observed at $L=2N$. Such
behavior of the correlation function is the result of using the
memory function of the step-wise form. To clarify this fact it is
convenient to change the variable $r$ by the band number $s$ and
the intra-band number $\rho$:
\begin{equation}\label{ro}
K(r)=K_s(\rho),\qquad r=sN+\rho+1,\quad \rho = 0,1, \ldots ,N-1,
\; s= 0, 1, \ldots .
\end{equation}
Within $s$th band, Eq.~(\ref{basem}) is the recurrence of the second
order with the term $\alpha \, K_{s-1}(\rho)$ that is determined at
the previous step, while finding the correlation function for the
$(s-1)$th band.

\subsubsection{General expression for the correlation function}
\label{GE}

In the zeroth band ($s=0,~1 \leqslant r \leqslant N$), as it was
shown above, the correlation function is constant,
\begin{equation}\label{sol0}
K_0(\rho)=K_0.
\end{equation}
For the first band ($s=1,~N+1 \leqslant r \leqslant 2N$), taking
into account that $K(r-N-1)=K_0 (\rho)$, we have
\begin{equation}\label{sol1}
K_1(\rho)=K_0\, (1-(1-\alpha N)(1+\alpha)^\rho ).
\end{equation}
The correlation function decreases quasi-continuously within the
first band. However, as it was mentioned above, there exists a
discontinuity in the $K(r)$ dependence at $r=N$. This
discontinuity disappears in the limiting case of the strong
persistence, $\alpha\rightarrow 1/N$.

Substituting the obtained formula~(\ref{sol1}) in
Eq.~(\ref{basem}), we find the solution $K_2(\rho)$ for the second
band ($s=2,~2N+1 \leqslant r \leqslant 3N$),
\begin{equation}\label{sol2}
K_2(\rho)=K_0\,(1-(1-\alpha N)((1+\alpha)^{\rho+N}-\rho \,\alpha
(1+\alpha)^{\rho-1}).
\end{equation}
The correlation function $ K(r)$ is continuous at the interface
between the first and second bands, $K_1(N)=K_2(0)$. However, its
first derivative of $K(r)$ is discontinuous here (see
Fig.~\ref{korf}). Using the induction method, one can easily derive
the formula for $K_s(\rho)$ in the $s$th band ($sN+1 \leqslant
r\leqslant (s+1)N$):
\begin{equation}\label{solN}
K_s(\rho)=K_0\,\left(1-(1-\alpha N)\sum\limits_{i=1}^{s}
(-\alpha)^{i-1} (1+\alpha)^{(s-i)N+\rho -i+1} \text
C_{(s-i)N+\rho}^{i-1}\right),
\end{equation}
\[
\quad C_n^k=\frac{\Gamma (n+1)}{\Gamma (k+1) \Gamma (n-k+1)}.
\]

It follows from Eq.~(\ref{solN}), that the first $(s-2)$ derivatives
of the correlation function $K(r)$ are continuous at the border
between the $(s-1)$th and $s$th bands, but the derivative of the
$(s-1)$th order  changes discontinuously. Under the condition
$\alpha N \ll 1$, Eq.~(\ref{solN}) takes a simpler form,
\begin{equation}\label{solM-00}
K_s(\rho)=K_0\alpha ^s \text C_{s+N-1-\rho}^{s}.
\end{equation}
It is seen that the correlation function decreases proportionally to
$\alpha ^s$ with an increase of the band number $s$.

It is not easy to analyze the asymptotical behavior of the function
$K(r)$ at large $s$ because the number of summands in
Eq.~(\ref{solN}) increases being proportional to $s$. It is the
reason to propose another approach for the asymptotical study of the
correlation function $K(r)$ at $s\gg 1$.

\subsection{Asymptotical study of the correlation function}
\label{AS}

\subsubsection{Derivation of the characteristic equation}
\label{ChE}

The general solution of linear recursion equations~(\ref{base}) can
be represented as the linear combination of $N$ different
exponential functions,
\begin{equation}\label{Gen-Sol}
K(r)=\sum\limits_{i=1}^{N}a_{i}\xi_{i}^r.
\end{equation}
To find the values of $\xi_{i}$, we substitute the fundamental
solution,
\begin{equation}\label{Char}
K(r)=\xi^r,
\end{equation}
into Eq.~(\ref{base}) and obtain the characteristic polynomial
equation of the order of $N$. Constant multipliers $a_{i}$ are to be
determined by initial conditions.

It is more convenient to use Eq.~(\ref{basem}) instead of
Eq.~(\ref{base}), that implies the characteristic equation of the
order of $N+1$,
\begin{equation}\label{solM-0}
\xi^{N+1}-(1+\alpha)\xi^N+\alpha=0.
\end{equation}
The extra root of this equation, $\xi=1$, appears as a consequence
of passing on to the equation of order of $N+1$ from that of the
order of $N$ . The corresponding coefficient, $a_i$, in
Eq.~(\ref{Gen-Sol}) is equal to zero because the correlation
function should decrease at $r\rightarrow \infty$.

\begin{figure}[h!]
\begin{centering}
\scalebox{0.48}[0.57]{\includegraphics{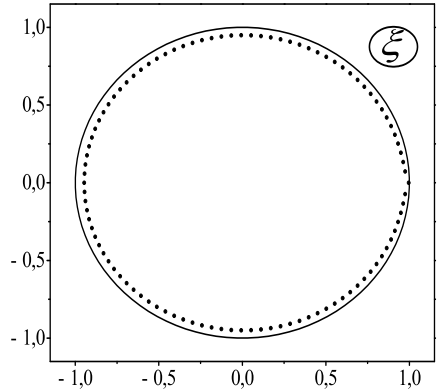}} \caption{The dots
are the roots of the characteristic equation~(\ref{solM-0}) for
$N=100$ and $\alpha=0.008$. The solid line is the circle $|\xi|=1$.}
\label{Root1}
\end{centering}
\end{figure}

Our study shows that Eq.~(\ref{solM-0}) has one real positive root
less than unity in the case of odd $N$. In the case of even $N$,
there are two real roots, one positive and one negative. The
remaining roots are complex. All absolute values of roots are less
than unity, that is in agreement with the finiteness of memory
function $F(r)$. In the case of large $N$, the absolute magnitudes
of all roots are close to unity for nearly all values of $\alpha$
satisfying the inequality,
\begin{equation}\label{con}
\frac{1}{N}\ln{\frac{1}{\alpha}}\ll 1.
\end{equation}
Distribution of the roots in the complex plane $\xi$ is shown in
Fig.~\ref{Root1}.

\protect

In the simplest case, $N=2$, Eq.~(\ref{solM-0}) has two real
roots,
\begin{equation}\label{root2}
\xi_{1,~2}=\frac{\alpha}{2}\pm \sqrt{\frac{\alpha^2}{4}+\alpha}.
\end{equation}
Taking into account the initial conditions, we find the solution of
Eq.~(\ref{base}) in the form,
\begin{equation}\label{sol30}
K(r)=\frac{\alpha}{4(1-\alpha)\sqrt{\alpha^2+4\alpha}}(\xi_1^{r-1}(1-\xi_2)-
\xi_2^{r-1}(1-\xi_1)).
\end{equation}
This expression can be simplified at small and large values of
parameter $\alpha$. For $\alpha\ll 1$, one obtains
\begin{equation}\label{0sol2}
K(r)=\frac{1}{4}\alpha^{[r/2]+1}
\end{equation}
with square brackets standing for the integer part. The
correlation function in the sequent odd and even points are equal
to each other. In accordance with Eq.~(\ref{solM-00}), $K(r)$
decreases at $r\rightarrow\infty$ being proportional to
$\alpha^s$. In the opposite limiting case of the strong
persistency, $\alpha\rightarrow 1/2$, we have two different roots,
\begin{equation}\label{rootp2}
\xi_1=1-\frac{4}{3}~\phi,\qquad
\xi_2=-\frac{1}{2}+\frac{1}{3}~\phi,
\end{equation}
with $\phi=1/2-\alpha$. The coefficient corresponding to the second
root is much less than that corresponding to the first one. Besides,
the second term in Eq.~(\ref{sol30}) decreases more rapidly.
Therefore, the approximate solution in this case is
\begin{equation}\label{1sol2}
K(r)=\frac{1}{4}\exp{-4\phi(r-1)/3}.
\end{equation}

\subsubsection{Correlation function at small $\alpha$} \label{TS}

Let us return to the case of arbitrary value of $N$. If $\alpha$
is very small, i.e. at
\begin{equation}\label{ñ2}
\frac{1}{N}\ln{\frac{1}{\alpha}}\gg 1,
\end{equation}
Eq.~(\ref{solM-0}) has $N$ roots with small absolute magnitudes:
\begin{equation}\label{0s}
\xi_k=\alpha^{1/N}(\cos(2\pi\frac{k}{N})+i\sin(2\pi\frac{k}{N})),\qquad
k=0,\ldots , N-1.
\end{equation}
The correlation function, being a linear combination of the power
functions with these roots as their exponents, decreases
proportionally to $\alpha^s$, which agrees with Eq.~(\ref{solM-00}).

The coefficients $a_i$ in the linear combination Eq.~(\ref{Gen-Sol})
can be found in a general case, without any restrictions imposed on
the value of $\alpha$. The solution of Eq.~(\ref{base}) written for
$1 \leqslant r \leqslant N-1 $ along with $K(0)=1/4$ can be
expressed with the help of the Vandermond determinants:
\begin{equation}\label{solz}
K(r)=K_0(\alpha N-1)\sum\limits_{k=1}^{N}\frac{\xi_k^{r-1}}
{\prod\limits_{j=1,j\neq k}^{N+1} (\xi_k-\xi_j)},
\end{equation}
with $\xi_{N+1}=1$.

\subsubsection{Correlation function at not too
small $\alpha$} \label{NTS}

In the case (\ref{con}) of not too small $\alpha$, the absolute
magnitudes of all roots are close to unity. It is convenient to
rewrite Eq.~(\ref{solM-0}), introducing two new real variables
$\gamma$ and $\varphi$ instead of complex $x$ according to
\begin{equation}\label{gamma}
x=(1-\frac{1}{\gamma N})e^{i \varphi}.
\end{equation}
Equation~(\ref{solM-0}) takes the form,
\begin{equation}\label{gammas}
\alpha (e^{\frac{1}{\gamma}-i N\varphi}-1)=(1-\frac{1}{\gamma
N})e^{i \varphi}-1.
\end{equation}
For the real root, Eq.~(\ref{solM-0}) yields,
\begin{equation}\label{gammas0}
\alpha N \gamma(e^{1/\gamma}-1)=1.
\end{equation}

This expression along with Eq.~(\ref{Char}) determines the
asymptotical behavior of correlation function. It was first obtained
in Ref.~\cite{uyakm}. The qualitative approach of this paper did not
allow one to take into account the contribution of other roots as it
is done in the present paper.

\begin{figure}[h!]
\begin{centering}
\scalebox{0.5}[0.43]{\includegraphics{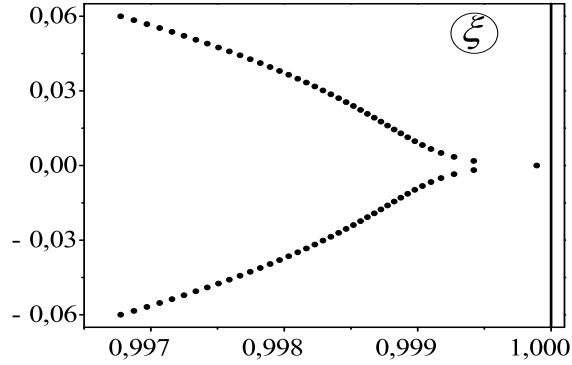}} \caption{The
roots of characteristic polynomial equation close to the point
$\xi=1$ for $N=4000$, $\alpha=2\cdot10^{-4}$. The solid line is
$\texttt{Re} \,\, \xi=1$.} \label{Root2}
\end{centering}
\end{figure}

Equation~(\ref{gammas}) yields all remaining complex roots with the
values of $\varphi$, which are quite uniformly distributed over the
circle $[0,2\pi]$ and
\begin{equation}\label{1/gam}
\frac{1}{\gamma}\sim \ln{\frac{1}{\alpha}}.
\end{equation}
The roots of Eq.~(\ref{solM-0}) located in the vicinity of point
$\xi=1$ are shown in Fig.~\ref{Root2}. The single real root is much
closer to the line $\texttt{Re}\,\, \xi=1$ than the other ones.
Besides, the coefficients $a_i$ in Eq.~(\ref{solz}) (see also
Eq.~(\ref{Gen-Sol})) for all terms  containing the complex exponents
are much less than those for the term with the real exponent.
Therefore, the behavior of correlation function $K(r)$ is determined
generally by the term with the real exponent.

\protect

The exact correlation function $K(r)$ resulting from the numerical
simulation of Eq.~(\ref{solz}) and its approximation determined by
the contribution of the real root alone are shown in
Fig.~\ref{korf}. These curves are compared with that obtained in
Ref.~\cite{uyakm} by a qualitative method.

\begin{figure}[h!]
\begin{centering}
\scalebox{0.6}[0.55]{\includegraphics{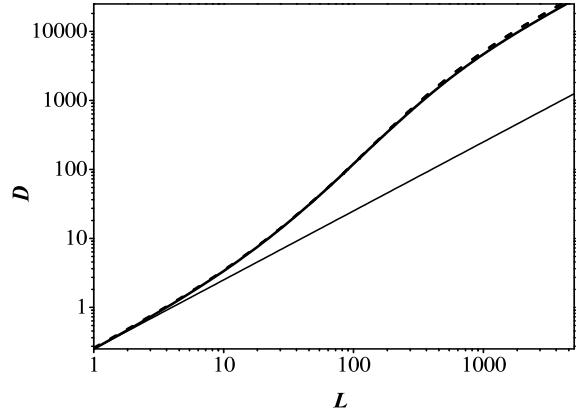}} \caption{Variance
$D(L)$ for the Markov chain with $N=100$, $\alpha=0.008$ obtained by
means of exact Eqs.~(\ref{solz}), (\ref{KD}) (solid line) and using
only one root of Eq.~(\ref{solM-0}) (dashed line). The thin solid
line describes the non-correlated Brownian diffusion, $D(L)=L/4$.}
\label{disp}
\end{centering}
\end{figure}

The obtained correlation function can be used to calculate one of
the most important characteristics of the random binary sequences,
the variance of number of unities in the $L$-word. The results of
the numerical simulations are shown in Fig.~\ref{disp}. One can see
a good agreement of curves plotted using both of these methods.

\protect

\subsubsection{Conclusion}

Thus, we have demonstrated the efficiency of describing the symbolic
sequences with long-range correlations in terms of the many-step
Markov chains with the \emph{additive} memory function. Actually,
the memory function appears to be a suitable informative "visiting
card" of any symbolic stochastic process. Various methods for
finding the memory function via the correlation function of the
system are proposed. Our preliminary consideration shows the
possibility to generalize our concept of the Markov chains on larger
class of random processes where the random variable can take on
arbitrary, finite or infinite number of values.

The suggested approach can be used for the analysis of different
correlated systems in various fields of science. For example, the
application of the Markov sequences to the theory of spin chains
with long-range interaction makes it possible to estimate some
thermodynamic characteristics of these non-extensive systems.


\begin{references}
\bibitem{ref} The existence of the "additional tail" in the correlation function
is in agreement with Ref.~\cite{uyakm} and corresponds to the well
known fact that the correlation length is always larger then the
region of memory function action.
\end{references}
\end{document}